\documentstyle[11pt,amstex,righttag]{article}
\addtocounter{secnumdepth}{1}
\newcommand{\dd}{\text{d}}
\newcommand{\ee}{\text{e}}
\newcommand{\ii}{\text{i}}

\numberwithin{equation}{section}

\begin{document}

\thispagestyle{empty}

\title
{Statistics of the one-dimensional
Riemann walk\\
}
\author{A.M. Mariz$^1$, F. van Wijland$^2$, H.J. Hilhorst$^2$,\\
 S.R. Gomes J\'unior$^1$, and C. Tsallis$^3$
\\[10mm]
 {\normalsize 
$^1$Departamento de F\'\i sica Te\'orica e Experimental}\\
 {\normalsize
Universidade Federal do Rio Grande do Norte}\\
 {\normalsize
Campus Universit\'ario,
59072-970 Natal, RN, Brazil}\\[2mm]
 {\normalsize
$^2$Laboratoire de Physique Th\'eorique$^*$}\\
 {\normalsize
B\^atiment 210, Universit\'e de Paris-Sud}\\
 {\normalsize
91405 Orsay Cedex, France}\\[2mm]
 {\normalsize
$^3$Centro Brasileiro de Pesquisas F\'\i sicas}\\
 {\normalsize
Rua Dr. Xavier Sigaud 150}\\
 {\normalsize
22290-180 Rio de Janeiro, RJ, Brazil}\\
}
\maketitle
\vspace{0cm}
\begin{small}
\begin{abstract}
\noindent The Riemann walk is the lattice version of the L\'evy flight.
For the one-dimensional Riemann walk of L\'evy exponent $0<\alpha<2$
we study the 
statistics of the {\it support}, {\it i.e.}
set of visited sites, after $t$ steps.
We consider a wide class of support related observables $M(t)$,
including the number $S(t)$
of visited sites and the number $I(t)$ 
of sequences of adjacent visited sites. 
For $t\to\infty$ we
obtain the asymptotic power laws for the averages, variances, and
correlations of these observables.
Logarithmic correction factors appear for $\alpha=\frac{2}{3}$ and
$\alpha=1$. {\it Bulk} and {\it
surface} observables 
have different power laws for $1\leq\alpha<2$. Fluctuations are shown to 
be universal
for $\frac{2}{3}\leq\alpha<2$. This means that in the limit $t\to\infty$
the deviations from
average $\Delta M(t)\equiv M(t)-\overline{M(t)}$
are fully described either by a single $M$ independent stochastic
process (when $\frac{2}{3}<\alpha\leq 1$) or by two such processes, one
for the bulk and one for the surface observables (when $1<\alpha<2$). \\

\end{abstract}
\end{small}
\vspace{15mm}

\noindent LPT ORSAY 00/23\\
{\small$^*$Laboratoire associ\'e au Centre National de la
Recherche Scientifique - UMR 8627}
\newpage
%%%%%%%%%%%%%%%%%%%%%%%%%%%%%%%%%%%%%%%%%
\section{Introduction} 
\label{secintroduction}

The L\'evy flight is a random walk in continuous space whose step size
distribution has a power law tail and is therefore sometimes called a
"L\'evy \cite{Levy} distribution".
The ubiquity of such distributions has been emphasized by many
authors, and is a consequence of the power law tail being invariant
under convolution. Many interesting
instances of the occurrence of L\'evy distributions are given by Tsallis
\cite{Tsallis} and  Tsallis {\it et al.} \cite{Tsallisetal}. These range
from applications in physics (superdiffusion, chaotic fluid flow) and
engineering (leaking taps)
through studies of the physiology of heart activity,
all the way to descriptions \cite{BouchaudPotters}
of fluctuations of financial markets. 

A one-dimensional
lattice version of the L\'evy flight may be constructed as follows. 
Let a random walk consist
of independent steps, and let the probability
$p(\ell)$ for a displacement of $\ell$ lattice units in a single step be given
by $p(0)=0$ and
\begin{equation}
p(\ell)=A\,|\ell|^{-1-\alpha} \qquad (\ell=\pm1,\pm2,\ldots)
\label{defRiemannwalk}
\end{equation}
Here $\alpha>0$ is the {\it L\'evy exponent}.
Normalization of $p$ implies that
$A^{-1}=2\,\zeta(1+\alpha)$
where $\zeta$ is the Riemann zeta function. 
This random walk was
first studied by Gillis and Weiss \cite{GillisWeiss} in 1970.
It is called the {\it Riemann walk} by Hughes (Ref.\,\cite{Hughes},
p.\,154) and we will conform to
that terminology. 
More generally we call {\it of Riemann type} any 
one-dimensional lattice
walk whose $p(\ell)$ is
asymptotically proportional to $|\ell|^{-1-\alpha}$ when $|\ell|\to\infty$.

Riemann type walks were 
reviewed in detail by Hughes \cite{Hughes}. 
Of particular interest is
the exponent regime $0<\alpha\leq 2$, where
these walks have a 
mean square displacement per step,
$\langle\ell^2\rangle$, which is infinite. 
There then exists, at least for
certain global walk features, a correspondence
between simple random walk on a $d$-dimensional lattice and
one-dimensional Riemann
type walks of exponent $\alpha=2/d$. In some ways the fraction
$\frac{2}{\alpha}$ acts as the walk's {\it
effective dimensionality.}
But whereas analytical results for noninteger dimension $d$ cannot be
checked by computer simulations, the full continuum of $\alpha$ values
is accessible to Monte Carlo studies.\\

Much interest has centered around
the following question. Let there be a $t$ step
Riemann walk. Then what are
the statistical properties of its {\it support} ${\cal S}(t)$, 
{\it i.e.,} of the set of sites that the walk has visited?
There appears immediately an
important difference between the exponent regimes
$0<\alpha<1$ and $1<\alpha<2$. In the former regime the Riemann walk is
transient \cite{Hughes,Weiss}
and it is easy to show (see Sec.\,\ref{secfractal}) that
${\cal S}(\infty)$ is a set of fractal dimension 
$d_{\cal S}=\alpha$. In the latter case the Riemann walk is recurrent
\cite{Hughes,Weiss},
${\cal S}(\infty)$ coincides with the full one-dimensional lattice,
and $d_{\cal S}=1$.   
The existing literature deals with the different question of finding 
the properties of ${\cal S}(t)$ for asymptotically $t$;
the results reflect, nevertheless, 
the same distinction between $0<\alpha<1$ and
$1<\alpha<2$. The borderline case $\alpha=1$ is more subtle. 

Gillis and Weiss \cite{GillisWeiss} 
study the number $S(t)$ of distinct sites in the support.
They find, among other results, that for $t\to\infty$ the average
of this random variable behaves \cite{GillisWeiss} as
$\overline{S(t)}\sim t$ for $0<\alpha<1$ and as $\overline{S(t)}\sim
t^{1/\alpha}$ for $1<\alpha<2$,
where $\,\sim\,$ indicates asymptotic proportionality. 
For $\alpha=1$ and $\alpha=2$ power laws with logarithmic correction
factors appear \cite{GillisWeiss}. 
For $\alpha>2$ the result $\overline{S(t)}\sim t^{1/2}$
is identical to that for the simple random walk in $d=1$.

A recent extension of this work is due to Berkolaiko {\it et al.}
\cite{Berkolaikoetal}. Pursuing a question initially asked for
the case of the simple random walk by Larralde {\it et al.}
\cite{Larraldeetal}, 
these authors investigate the number $S_N(t)$ of distinct
sites visited by $N$ independent $t$ step Riemann type walks all
starting on the same lattice site. Again power laws appear, both for
$t\to\infty$ at fixed $N$ and for $N\to\infty$ at fixed $t$.\\

The present work extends the investigations of Gillis and Weiss into
a different direction. 
We limit ourselves to the Riemann walk defined by
Eq.\,(\ref{defRiemannwalk}), with
$\alpha$ in the regime of greatest interest, that is, $0<\alpha<2$.
Our results may be summarized under three headings.

{\it 1.\, Variance} $\,\overline{\Delta S^2(t)}$.\,\, 
For any quantity $X(t)$ we will denote
its instantaneous deviation from average by $\Delta X(t)\equiv
X(t)-\overline{X(t)}$.
Traditionally in this field the calculation of the average number
$\overline{S(t)}$ of distinct sites visited has
been followed by a calculation of the variance $\overline{\Delta
S^2(t)}$
of that number. Thus, for the simple random walk $\overline{S(t)}$
was first calculated by Dvoretzky and Erd\"os \cite{DvoretzkyErdos} in
1951, and $\overline{\Delta S^2(t)}$ by Jain and Pruitt \cite{JainPruitt}
in 1970. For the one-dimensional Riemann walk the present work
supplements
the 1970 results due to Gillis and Weiss \cite{GillisWeiss} for
$\overline{S(t)}$ by the corresponding ones for the variances
$\overline{\Delta S^2(t)}$ in the regime $0<\alpha<2$.

{\it 2.\, Variables other than $S(t)$.}\,\, 
We use the powerful
generating function method (GFM), which was introduced into the field of
random walks by Montroll \cite{Montroll}
and Montroll and Weiss \cite{MontrollWeiss}. Overviews of this method
are given by Weiss \cite{Weiss} and by Hughes \cite{Hughes}. The
first calculation of a variance by 
the GFM, {\it viz.} that of $S(t)$
for the simple random walk, is due to Torney 
\cite{Torney} in 1986.

In 1994 Coutinho {\it et al.}~\cite{Coutinhoetal}
performed Monte Carlo simulations of, among other things, the number
of unvisited islands enclosed by the support of the
$t$ step simple random walk in two dimensions.
This led Caser and Hilhorst \cite{CaserHilhorst} to 
analytically determine the asymptotic behavior of the average number of
islands. Subsequently Van Wijland {\it et al.} \cite{WCH,WH}
developed
a compact GFM based analytical scheme for calculating
simultaneously the averages, variances, and correlations of a large
class of observables characteristic of the support, 
generically denoted by the symbol $M(t)$. In $d=2$ this class includes
also the total boundary length of the support, and in
$d=3$ its surface area and Euler index.

Here we bring this scheme to bear on the one-dimensional
Riemann walk. 
The support of this walk consists of alternating sequences of visited
and unvisited sites. Among the most prominent members of the
class of observables $M(t)$  is, next to $S(t)$,
the number of visited sequences, that we will denote by $I(t)$. Islands
in $d=1$ just are unvisited sequences enclosed by the support, of which
there are $I(t)\!-\!1$; the support furthermore has $2I(t)$ boundary sites
(= visited sites adjacent to an unvisited one).
Table I summarizes our results for the asymptotic laws of the
averages, variances, and correlations involving $S(t)$ and $I(t)$.
Beyond their intrinsic interest these laws may serve in heuristic
arguments in reaction--diffusion processes, {\it e.g.}, 
to estimate the trapping probability of an atom
that diffuses in a random absorbing environment, or the effective
reaction rate between two diffusing species.
We defer further comments to Sec.\,\ref{secconclusions}.
\begin{table}[tbh]
\begin{center}
\begin{tabular}{||c|c|c|c|c|c||}
\hline
&$0\!<\!\alpha\!<\!\frac{2}{3}$&
$\alpha=\frac{2}{3}$&
$\frac{2}{3}\!<\!\alpha\!<\!1$&$\alpha=1$&$1\!<\!\alpha\!<\!2$\\
\hline
$\begin{array}{l}
\overline{S(t)}\\
\overline{I(t)}
\end{array}
$
&$t$&$t$&$t$&
$\begin{array}{l}
t\,\log^{-1}t\\
t\,\log^{-2}t
\end{array}
$
&
$\begin{array}{l}
t^{\frac{1}{\alpha}\phantom{-1}}\\
t^{\frac{2}{\alpha}-1}
\end{array}
$
\\
\hline
$\overline{\Delta S^2(t)}$&&&&
$t^2\log^{-4}t$&$t^{\frac{2}{\alpha}\phantom{-1}}$\\
$\overline{\Delta S(t)\Delta I(t)}$&$t$&
$t\log t$&$t^{4-\frac{2}{\alpha}}$&
$t^2\log^{-5}t$&$t^{\frac{3}{\alpha}-1}$\\
$\overline{\Delta I^2(t)}$&&&&
$t^2\log^{-6}t$&$t^{\frac{4}{\alpha}-2}$\\
\hline
\end{tabular}
\end{center}
\noindent Table I.
{\small 
\; Leading asymptotic behavior 
as $t\to\infty$ of the averages, variances, and
correlation of $S(t)$ and $I(t)$ in different regimes of the L\'evy
exponent $\alpha$. The exact prefactors of the asymptotic laws
are given in the text.
The results for $\overline{S(t)}$ are due to Gillis and
Weiss \cite{GillisWeiss}; the result for $\overline{\Delta S^2(t)}$ 
in the range $\,0<\alpha<\frac{2}{3}\,$ follows from the theorem of
Jain, Orey, and Pruitt (see Hughes \cite{Hughes}, p.\,344); 
all others are new.} 
\end{table}

{\it 3.\, Universality of fluctuations.}\,\, 
The deviations from average $\Delta S(t)$,$\Delta I(t)$,
$\ldots$,$\Delta M(t)$,$\ldots$ are randomly time-dependent
variables that one would {\it a priori}
expect to exhibit some degree of correlation. One calls these
fluctuations {\it universal} -- by lack of a better name -- when in the
limit $t\to\infty$ all $\Delta M(t)$ are 
asymptotically equal (up to a proportionality constant) to a {\it
single} $M$ independent stochastic process.
For the simple random walk universality was shown to hold in dimensions
$d=2$ \cite{WCH} and $d=3$ \cite{WH}, and not to hold in
$d=4,5,\ldots$. For the $d=1$ Riemann walk we find that universality
holds in the exponent regime $\frac{2}{3}\leq\alpha<2$, but not for
$0<\alpha<\frac{2}{3}$. A novelty with respect to the case of
the simple random walk is that for $1<\alpha<2$ not a single,
but {\it two} $M$ independent processes are needed to describe the
universal fluctuations: one applies to bulk and the other to surface
observables. 
The precise statements are given in Sec.\,\ref{secuniversality}.\\

This article is set up as follows.
Sec.\,\ref{secmethod} describes those elements of
our analysis that are common to
the full exponent interval $0<\alpha<2$.
Secs.\,\ref{secalpha01} and \ref{secalpha12} 
deal more in particular with the exponent regimes
$0<\alpha<1$ and $1<\alpha<2$, respectively, and derive the asymptotic
behavior of averages, variances, and correlations. In Sec.\,\ref{secspecial}
we do the same for the exceptional values $\alpha=\frac{2}{3}$ and
$\alpha=1$. In Sec.\,\ref{secuniversality} we discuss the universality
properties. 
In Sec.\,\ref{secconclusions} we provide some additional interpretation of
our results and conclude.

\section{Observables, averages, and correlations}
\label{secmethod}

\subsection{Observables $M(t)$}
\label{secobservables}

Our analysis is based on first writing
quantities of interest in terms of the field $m(x,t)$ of "complementary
occupation numbers" defined by
$m(x,t)=1$ if site $x$ has not yet been visited at time $t$, and
$m(x,t)=0$ otherwise.
The expressions of $S$ and $I$ in terms of $m$ are
\begin{eqnarray}
S(t)&=&\sum_{x=-\infty}^{\infty}\,[1-m(x,t)]
\label{Im}\\
I(t)&=&\sum_{x=-\infty}^{\infty}\,m(x,t)[1-m(x+1,t)]
\label{Sm}
\end{eqnarray}
$S$ and $I$ are representatives of a 
general class of "observables" $M$
that are sums on $x$ of a summand to which each lattice site contributes
a factor $m$\, $1-m$, or $1$, {\it i.e.,} the summand tests for the
presence of a specific pattern of visited ({\it "black"}) and unvisited
({\it "white"}) sites. The following slightly more abstract
characterization of the $M$ will be needed.
Let $A=\{a\}$ be a finite set of distinct nonnegative integers $a$,
such that either $A=\emptyset$ or, if not, $A$ includes the element $a=0$.
The general observable $M(t)$ that we will consider is
\begin{equation}
M(t)=\sum_{x=-\infty}^{\infty}\sum_A\mu_A\prod_{a\in A}m(x+a,t)
\label{defM}
\end{equation}
where for $A=\emptyset$ the product is equal to unity and where
$\{\mu_A\}$  
is a set of numerical coefficients characteristic of $M$.
When their $M$ dependence needs to be indicated we will write $\mu_A[M]$.
Eqs.\,(\ref{Sm}) and (\ref{Im}) show that
$S(t)$ and $I(t)$ are of the form of
Eq.\,(\ref{defM}) with only two nonzero coefficients, as shown in 
Table II.
\begin{table}[tbh]
\begin{center}
\begin{tabular}{||l|r r r r||}
\hline
\phantom{\{}$A$                     &$\mu_A[S]$&$\mu_A[I]$&$\mu_A[S_1]$&$\mu_A[I_1]$\\
\hline
$\phantom{\{} \emptyset$&  1 &    &    &$  $\\
\{0\}                   &$-1$&  1 &    &$ 1$\\
\{0,\,1\}               &    &$-1$&    &$-2$\\
\{0,\,2\}               &    &    &$ 1$&    \\
\{0,\,1,\,2\}           &    &    &$-1$&$ 1$\\
\hline
\end{tabular}
\end{center}

\noindent Table II.
{\small
\; Coefficients $\mu_A[M]$ for the four observables $M=S, I, S_1,I_1$
defined in the text; entries not shown are zero. 
The coefficients in each column add up to zero.}
\end{table}

Two further examples of observables of type (\ref{defM}) are the total number
$S_1(t)$ of visited sequences consisting of only a single site, and the
total number $I_1(t)$ of single-site unvisited sequences. Their 
coefficients $\mu_A$ involve sets $A$ of up to three elements;
they are easily determined and have also been listed
in Table~II.

The following remarks, important for later, are verified without much
effort. The coefficient $A_\emptyset$ is nonzero if and only if $M$ is
built up exclusively out of factors $m$. Since these correspond to
visited sites, that make up the "bulk" of the support,
we will call an $M$ of this type a {\it bulk}
observable. Observables built up exclusively out of factors $m$ do not
occur, since their expectation value on an infinite lattice is
infinite. Hence the remaining observables refer to patterns consisting
of both visited and
unvisited sites, and we will therefore call them {\it surface} observables. 
[In the terminology of Refs.\,\cite{WCH,WH} these are {\it "black"}
and {\it "black-and-white"} observables. They might also be called {\it
"S-like"} and {\it "I-like"}, respectively.]
The distinction between these two subclasses will play a role 
only in the exponent regime $1\leq\alpha<2$.

\subsection{Basic formulas for averages and correlations}
\label{secgeneralities}

In this work we will first 
evaluate the $t\to\infty$ behavior  
of the averages $\overline{M(t)}$. Then we turn to 
the covariance matrix $\overline{\Delta M(t)\Delta M'(t)}$,
where $M'(t)$ is a second observable with coefficients $\mu'_A$.
Although the authors of Refs.\,\cite{WCH} and \cite{WH} 
deal with the simple random walk, the larger part of their formal
developments also holds for the Riemann walk. 

The averages $\overline{M(t)}$ and $\overline{M(t)M'(t)}$
can be obtained as follows
\cite{WCH,WH}.
Let $G(x,t)$ be the Green function of the
one-dimensional Riemann walk, that is,
the probability for a walker starting at the
origin to occupy site $x$ after $t$ steps. Let
$\hat{G}(x,z)=\sum_{t=0}^\infty z^t G(x,t)$ denote its
generating function 
and let 
{\bf G}$_A(z)$ be the $|A|\times|A|$ matrix of elements $\hat{G}(a-a',z)$
with $a,\,a'\in A$.
From this matrix one constructs the "inverse sum"
${\sf G}_A(z)$ defined by
\begin{equation}
{\sf G}_A^{-1}(z) = \sum_{a,a'\in A}[\text{\bf G}_A^{-1}(z)]_{aa'}
\label{defGA}
\end{equation}
These scalars satisfy certain elementary relations stated in Appendix A
as {\sc Properties 1--3}.
Two functions $C_M(z)$ and
$C_{MM'}(z)$ are defined in terms of the ${\sf G}_A(z)$ according to
\begin{eqnarray}
C_M(z)&=&\sum_{A\neq\emptyset}\,\mu_A\,\frac{1}{{\sf G}_A(z)} \label{defCM}\\
C_{MM'}(z)&=&
\sum_{A\neq\emptyset}\sum_{B\neq\emptyset}
\mu_A\mu'_B
\sum_{r=-\infty}^\infty\Big[ \frac{1}{{\sf G}_{A\cup(r+B)}(z)} 
-\frac{1}{{\sf G}_A(z)} -\frac{1}{{\sf G}_B(z)} \Big]\phantom{xx}
\label{defCMM}
\end{eqnarray}
Here $A\cup(r+B)$ denotes the union of the set $B$, translated by $r$,
and $A$.
The averages $\overline{M(t)}$ and  
$\overline{M(t)M'(t)}$ are then obtained as \cite{WCH,WH}
\begin{eqnarray}
\overline{M(t)}&=&-\frac{1}{2\pi\ii}\oint\frac{\dd z}{z^{t+1}}
\frac{1}{(1-z)^2}\,C_M(z) \label{defMav}\\
\overline{M(t)M'(t)}&=&
-\frac{1}{2\pi\ii} \oint\frac{\dd z}{z^{t+1}}
\frac{1}{(1-z)^2}\,C_{MM'}(z)
\label{defMMav}
\end{eqnarray}
where the integrations are counterclockwise around the origin.

The identities (\ref{defGA})--(\ref{defMMav}) are fundamental to
random walk theory; they hold for any translationally invariant random
walk, whether on a finite lattice with periodic boundary conditions or
on an infinite lattice.
They allow for the calculation, in a very compact way, of many known and
new results. 

{\it Special cases.}\, When $M(t)=S(t)$,
the following simplifications occur. The sums on $A$ and on $B$
in Eqs.\,(\ref{defCM}) and (\ref{defCMM}) then have only the
single term with $A=\{0\}$ and $B=\{0\}$,
respectively. 
Furthermore {\sf G}$_{\{0\}}(z)=\hat{G}(0,z)$, the matrix {\bf
G}$_{\{0\}\cup(r+\{0\})}(z)$ is two by two, and an easy calculation leads to
{\sf G}$_{\{0\}\cup(r+\{0\})}(z)=\frac{1}{2}(\hat{G}(0,z)+\hat{G}(r,z))$.
When $M(t)=I(t)$, the sum in Eq.\,(\ref{defCM}) involves {\sf
G}$_{\{0\}}(z)$ and {\sf G}$_{\{0,1\}}(z)
=\frac{1}{2}(\hat{G}(0,z)+\hat{G}(1,z))$.
The sums on $A$ and $B$ in Eq.\,(\ref{defCMM})
then lead to four terms, which may be evaluated with a little more effort.

\subsection{Limit $\,t\to\infty$ and scaling limit}
\label{limit}

Explicit evaluation of the general expressions
(\ref{defGA})
-(\ref{defMMav})
is limited in practice by the
calculation of the inverse sums
${\sf G}_A$, which require the inversion of a
matrix of dimension $|A|$.
Similarly, evaluation of ${\sf G}_{A\cup(r+B)}$ 
is an inversion problem
of dimension $|A|+|B|$ (when 
$A$ and $r+B$ have an empty intersection). It turns out that the sum
on $r$ in Eq.\,(\ref{defCMM}) can be performed only 
%after ${\sf G}_{A\cup(r+B)}$ 
%has been worked out more explicitly. It turns out that this can be done 
in the scaling limit
\begin{equation}
z\to 1,\quad  |r|\to\infty \qquad {\mbox{with}} \quad
\xi= r(1-z)^{\frac{1}{\alpha}} 
\quad {\mbox{fixed}}
\label{sclimit}
\end{equation}
Finally, it will be possible to evaluate
the integrals in Eqs.\,(\ref{defMav}) and (\ref{defMMav}) 
only asymptotically for
$t\to\infty$, a limit already implied by Eq.\,(\ref{sclimit}).

In order to prepare for these limits we rewrite the
preceding expressions as follows.
Using the simplified notation $G_0(z)=\hat{G}(0,z)$ we split
the generating function $\hat{G}(x,z)$ up according to
\begin{equation}
\hat{G}(x,z)=G_0(z)-g(x,z)
\label{defg}
\end{equation}
In full analogy to {\bf G}$_A(z)$ we define {\bf g}$_A(z)$
as the matrix of elements $g(a-a',z)$ with $a,\,a'\in A$,
and ${g}_A^{-1}$ as the sum of all elements of {\bf g}$_A^{-1}$.
Let now {\bf J} be the square matrix of elements $J_{aa'}=1$.
Then 
\begin{equation}
{\text{\bf G}}_A(z) = G_0(z)\,{\text{\bf J}}\, -\, {\text{\bf g}}_A(z)
\end{equation}
and, by {\sc Property 1} of Appendix A,
\begin{equation}
{\sf G}_A(z)=G_0(z)-{g}_A(z)
\label{relGg}
\end{equation}
This splitup will be useful for studying
the $z\to 1$ behavior of ${\sf G}_A(z)$. 
Although $G_0(z)$ may $(1\leq\alpha<2)$ or may not $(0<\alpha<1)$ diverge
as $z\to 1$, the functions $g(x,z)$ and $g_A(z)$ remain finite in that
limit. 

We now turn to the inverse sum
${\sf G}_{A\cup(r+B)}$ constructed from the matrix
{\bf G}$_{A\cup(r+B)}$. The dimension of this matrix is typically $|A|+|B|$.
Let {\bf J}$^{AB}$ be the $|A|\times |B|$ matrix
with all $J^{AB}_{ab}=1$. 
We then have (for $A\cap B=\emptyset$)
\begin{equation}
{\text{\bf G}}_{A\cup(r+B)}(z)=
\left(
\begin{array}{cc}
{\text{\bf G}}_A(z) & \hat{G}(r,z){\text{\bf J}}^{AB}\\[1mm]
\hat{G}(r,z){\text{\bf J}}^{BA} & {\text{\bf G}}_B(z)
\end{array}
\right) +
\left(
\begin{array}{cc}
0 & {\text{\bf V}}\\[1mm]
{\text{\bf V}}^T & 0
\end{array}
\right)
\label{splitGAB}
\end{equation}
where {\bf V} is the matrix of elements
\begin{equation}
V_{a,r+b}=\hat{G}(r+b-a,z)-\hat{G}(r,z)
\qquad a\in A,\,\,b\in B
\label{defV}
\end{equation}
and {\bf V}$^T$ is its transpose. 
The first matrix on the RHS of (\ref{splitGAB}) has the form (\ref{Lprop3}) of
Appendix A.
Applying {\sc Property 3} to that matrix we conclude that
\begin{equation}
\frac{1}{{\sf G}_{A\cup(r+B)}(z)} \,=\,
\frac{{\sf G}_A(z)+{\sf G}_B(z)-2\hat{G}(r,z)}{{\sf G}_A(z){\sf G}_B(z)-\hat{G}^2(r,z)}
\,\,+\,\,{\cal O}({\text{\bf V}}^2)
\label{decompGAB}
\end{equation}
where we anticipate, and will have to show later, that {\bf V} is small,
that the correction terms are of order
${\cal O}({\bf V}^2)$, and that they are negligible for our purpose.

Further analysis depends on the exponent $\alpha$.
We consider the two main regimes
$0<\alpha<1$ and
$1<\alpha<2$
in Secs.\,\ref{secalpha01} and \ref{secalpha12}, respectively.
The exceptional values 
$\alpha=\frac{2}{3}$ and $\alpha=1$ are 
discussed in Sec.\,\ref{secspecial}.\\

\subsection{Riemann walk Green function}
\label{green}

All quantities of interest have been expressed above
in terms of the Riemann walk Green function $\hat{G}(x,z)$. 
An elementary calculation yields
\begin{eqnarray}
\hat{G}(x,z)&=&\int_{-\pi}^\pi\frac{\dd q}{2\pi} 
\,\frac{\ee^{-\ii q x}}
{1-z\lambda(q)}\label{Gxz}\\
\lambda(q)&=&\frac{1}{\zeta(1+\alpha)}\,\sum_{\ell=1}^\infty
\ell^{-1-\alpha}\,\cos\ell q
\label{lambdaq}
\end{eqnarray}
The $q\to 0$ behavior of $\lambda(q)$ is crucial for the large scale
features of the Riemann walk. It is known \cite{GillisWeiss,Hughes} that
\begin{equation}\label{lambdaqpetit}
\lambda(q) = 1-C_\alpha\, |q|^\alpha + {\cal O}(q^2)  
\qquad (q\to 0)
\end{equation}
where for completeness we state the explicit expression
\begin{equation}
C_\alpha^{-1} = 2\zeta(1+\alpha)\,
\Gamma(1+\alpha)\,/\,[\pi\sin(\alpha\pi/2)]
\end{equation}
One finds by standard methods (see {\it e.g.} \cite{Hughes}) that in 
the limit $z\rightarrow 1$ the Green function in the origin $G_0(z)$
has the asymptotic expansion
\begin{eqnarray}
G_0(z)&=&G_0(1)-B_{\alpha}(1-z)^{\frac{1}{\alpha}-1}+{\cal O}(1-z)
\quad(0\!<\!\alpha\!<\!1;\alpha\neq\mbox{$\frac{1}{2}$})\phantom{xx}
\label{G2}\\[2mm]
G_0(z)&=&\mbox{$\frac{1}{3}$}\log [c(1-z)^{-1}]+{\cal O}(1-z)
\qquad(\alpha=1)
\label{G2.5}\\
G_0(z)&=&A_\alpha (1-z)^{-1+\frac{1}{\alpha}} + {\cal O}(1)
\qquad(1<\alpha<2)
\label{G3}
\end{eqnarray}
where $B_\alpha$ and $A_\alpha$ are the constants
\begin{eqnarray}
B_\alpha&=&-C_\alpha^{1/\alpha}/[2\sin(\pi/\alpha)]
\qquad(\mbox{$\frac{1}{2}$}<\alpha<1)
\label{defB}\\
A_\alpha&=&1/[2\alpha\,
C_\alpha^{1/\alpha}\sin(\pi/\alpha)]\qquad(1<\alpha<2)
\label{defA}
\end{eqnarray}
and $c$ is a constant such that there is no ${\cal O}(1)$ term in
Eq.\,(\ref{G2.5}). 
In Eqs.\,(\ref{G2})--(\ref{G3}) 
and elsewhere we use the following convention. The 
symbol $\,{\cal O}(X)\,$ 
indicates terms that are {\it
of order X\,} in the applicable limit ($X\to 0$ or $X\to\infty$); this
however is not to say that all preceding terms are larger. Thus,
the nonanalytic term in Eq.\,(\ref{G2}) is larger than
the ${\cal O}(1-z)$ terms only for $\frac{1}{2}\!<\!\alpha\!<\!1$. 
For $0\!<\!\alpha\!<\!\frac{1}{2}$ it is present only as a correction to the ${\cal
O}(1-z)$ terms;
the expression for its coefficient $B_\alpha$ in that regime
is different from  Eq.\,(\ref{defB}) but
will not be needed. For the borderline case $\alpha=\frac{1}{2}$,
excluded from Eq.\,(\ref{G2}), we have
$G_0(z)=G_0(1)-\frac{2}{\pi}(1-z)\log(1-z)^{-1}+{\cal O}(1-z)$;
but this special nonanalytic behavior will stay subdominant everywhere in
the remainder. 

From Eqs.~(\ref{Gxz}) and (\ref{lambdaqpetit}) one deduces that in the 
scaling limit (\ref{sclimit})
\begin{equation}
\hat{G}(r,z)\simeq (1-z)^{\frac{1}{\alpha}-1}F(\xi)\qquad(0<\alpha<2)
\label{Grzscaling}
\end{equation}
where $\xi=r(1-z)^{\frac{1}{\alpha}}$ and $F(\xi)$ is the scaling function 
\begin{equation}
F(\xi)=\int_{-\infty}^{\infty}\frac{\dd k}{2\pi}\,\,\frac{\ee^{-\ii
k\xi}}{1+C_\alpha |k|^\alpha}
\label{defF}
\end{equation}
For $\,\xi\to 0\,$ it behaves as
\begin{eqnarray}
F(\xi) &\simeq& 
2\alpha\,\zeta(1+\alpha)/[\pi\sin(\alpha\pi)]\,\,\xi^{-1+\alpha}
\quad (0<\alpha<1)
\label{xito01}\\[2mm]
F(\xi) &\simeq& {\mbox{$\frac{1}{3}$}}
\log\xi^{-1} \qquad(\alpha=1)
\label{xito01.5}\\[2mm]
F(\xi)&=&
A_\alpha+{\cal O}(\xi^{\alpha-1}) \quad\qquad (1\leq\alpha<2)
\label{xito02}
\end{eqnarray}
In Secs.\,4 and 5 we will also use the function $f(r,z)$ defined by
\begin{equation}
\hat{G}(r,z)=G_0(z)f(r,z)\quad\qquad (1\leq\alpha<2)
\label{deff}
\end{equation}
In the scaling limit one has
$f(r,z)\simeq f(\xi)=F(\xi)/F(0)$ when $1<\alpha<2$.

\subsection{Support at $t=\infty$}
\label{secfractal}

Whereas the remainder of this paper deals with the large $t$ behavior,
we briefly comment here on the structure of the support ${\cal S}(t)$
{\it at} $t=\infty$. 

As is well-known \cite{Weiss,Hughes}), 
random walks are recurrent (are transient) if
$G_0(1)=\infty$\, (if $G_0(1)<\infty$).
The Riemann walk of this work
is recurrent for $1\leq\alpha<2$, which means that all
sites are visited with probability 1, and that {\it at}
$t=\infty$ the support ${\cal S}(\infty)$
coincides with the full one-dimensional
lattice.

For $0<\alpha<1$, however, the Riemann walk is transient, so that at 
$t=\infty$ the support ${\cal S}(\infty)$
will still be only a subset of the full lattice.
We may estimate the average number of visited sites $\Sigma_{L}$ 
between $x=-L$ and $x=L$ in ${\cal S}(\infty)$.
According to standard random walk theory \cite{Weiss,Hughes}
\begin{equation}
\Sigma_{L}=\sum_{x=-L}^L \frac{\hat{G}(x,1)}{G_0(1)}
\label{fractal}
\end{equation}
Upon substituting (\ref{Gxz}) in (\ref{fractal}) one easily evaluates 
$\Sigma_L$ for asymptotically large $L$, with the result that
$\Sigma_L\sim L^{\alpha}$. It follows that the support ${\cal S}(\infty)$
has fractal dimension $d_{\cal S}=\alpha$.

%%%%%%%%%%%%%%%%%%%%%%%%%%%%%%%%%%%%%%%%%%%%%%%%%%%%%%%%%
\section{Riemann walk of exponent $\,0<\alpha<1$}
\label{secalpha01}

\subsection{Averages}
The large time behavior of $\overline{M(t)}$ comes from the 
behavior of $C_M(z)$, defined in Eq.~(\ref{defCM}),
in the limit $z\to 1$.  
From Eq.\,(\ref{defg})
and the explicit expressions (\ref{Gxz}) and (\ref{lambdaq})
it may be shown that $g(x,z)=g(x,1)+{\cal O}(1-z)$
for all $0<\alpha<1$. Hence
\begin{equation}
{g}_A(z)={g}_A(1)+{\cal O}(1-z)
\label{expgA}
\end{equation}
after which it follows from 
Eqs.~(\ref{relGg}), (\ref{G3}), and (\ref{expgA}) that
\begin{equation}
{\sf G}_A(z)={\sf G}_A(1)-B_\alpha(1-z)^{\frac{1}{\alpha}-1}+{\cal O}(1-z)
\label{expGA}
\end{equation}
Inverting this relation and substituting in Eq.\,(\ref{defCM}) gives
\begin{equation}
C_M(z)=-m_1-
B_\alpha m_2(1-z)^{\frac{1}{\alpha}-1}-
B_\alpha^{2}m_3(1-z)^{\frac{2}{\alpha}-2}+\ldots+{\cal O}(1-z)
\label{exprCM}
\end{equation}
where the $m_n$ are determined by the coefficients $\mu_A$
of the observable $M$ according to
\begin{equation}
m_n[M]=\sum_{A\neq\emptyset}\frac{\mu_A}{{\sf
G}_A^n(1)}\qquad(n=1,2,\ldots;\,\,\,0<\alpha<1)
\label{defmnM}
\end{equation}
In Eq.\,(\ref{exprCM}) the dots stand for a power series in
$(1-z)^{\frac{1}{\alpha}-1}$ and 
the number of nonanalytic terms 
between the zeroth and the first power of $1-z$ is equal to
$n_\alpha\equiv
\lceil\frac{\alpha}{1-\alpha}\rceil-1$. That is, $n_\alpha$ is zero
for $0<\alpha<\frac{1}{2}$ and, as $\alpha$ goes up, 
jumps to $1,2,3,\ldots$ at $\alpha=
\frac{1}{2},\frac{2}{3},\frac{3}{4},\ldots,$ respectively. 
After Laplace inversion we get for the average
$\overline{M(t)}$ in the limit $t\to\infty$ the explicit result
\begin{equation}
\label{moyenneM}
\overline{M(t)}=m_1t+\frac{B_\alpha}{\Gamma(3-\frac{1}{\alpha})}m_2
t^{2-\frac{1}{\alpha}}+\frac{B_\alpha^2}{\Gamma(4-\frac{2}{\alpha})}m_3
t^{3-\frac{2}{\alpha}}+\ldots
+{\cal O}(1)
\label{threefive}
\end{equation}
where the number of nonanalytic terms between the leading and the ${\cal
O}(1)$ term is again equal to $n_\alpha$.

\subsection{Correlations}
\label{seccorrelations1}

In this subsection we consider two -- possibly equal -- 
observables $M$ and $M'$,
represented by sets of coefficients $\{\mu_A\}$ and $\{\mu'_A\}$,
respectively, and wish to study their correlation.
The starting point is 
Eq.~(\ref{defCMM}) for $C_{MM'}(z)$, in which we substitute 
Eq.\,(\ref{decompGAB}). 
Whereas ${\sf G}_A(z)$ and ${\sf G}_B(z)$ tend to finite values
in the limit $z\to 1$, the Green function $\hat{G}(r,z)$ vanishes in that
limit when taken with $\xi$ fixed. This suggests that we expand
in powers of $\hat{G}(r,z)$,
\begin{equation}
\frac{1}{{\sf G}_{A\cup(r+B)}(z)}-\frac{1}{{\sf G}_A(z)}-\frac{1}{{\sf G}_B(z)}=\sum_{n=1}^{\infty}C_{AB}^{(n)}(z){\hat{G}}^n(r,z)
\,+\,{\cal O}({\text{\bf V}}^2)
\label{expGAB}
\end{equation}
with coefficients
\begin{equation}
C_{AB}^{(n)}(z)=\left\{
\begin{array}{ll}
[{\sf G}_A(z)+{\sf G}_B(z)]
[{\sf G}_A(z){\sf G}_B(z)]^{-\frac{n}{2}-1}&(n\text{ even})\\[4mm]
-2\,[{\sf G}_A(z){\sf G}_B(z)]^{-\frac{n+1}{2}}&(n\text{ odd})
\end{array}\right.
\label{Cn}
\end{equation}
The sum over space that occurs in Eq.\,(\ref{defCMM}) 
leads us to now consider the sums $\sum_r
{\hat{G}}^n(r,z)$.
From conservation of probability one finds that for $n=1$
\begin{equation}\label{fred1}
\sum_r\hat{G}(r,z)=(1-z)^{-1}
\label{defFalpha1}
\end{equation}
For general $n$ the calculation of $\sum_r
{\hat{G}}^n(r,z)$ is slightly more laborious; after substituting
Eq.\,(\ref{Gxz}) for $\hat{G}$
one finds by explicit expansion in powers of $1-z$
that for $z\to 1$ the sum on $r$ behaves as
\begin{equation}\label{sumGn}
\sum_{r}{\hat{G}}^n(r,z)\simeq F_{\alpha,n}(1-z)^
{-1+(n-1)(\frac{1}{\alpha}-1)}+{\cal O}(1)
\qquad(0\!<\!\alpha\!<\!1;\alpha\neq 1-\mbox{$\frac{1}{n}$})
\label{sumG3}
\end{equation}
For $n=1-\frac{1}{1-\alpha}$ (with $n=1,2,\ldots$) the sum on $r$ instead
behaves as $\log(1-z)$; this happens, in particular, for $n=3$ when
$\alpha=\frac{2}{3}$, a case studied separately in Sec.\,\ref{secspecial}. 

We will now continue to consider the generic case.
The nonanalytic term on the RHS of Eq.\,(\ref{sumG3}) dominates
the ${\cal O}(1)$ term only for 
$n<\frac{1}{1-\alpha}$. In that case (\ref{sumG3}) follows just from
the scaling form (\ref{G3}) of $\hat{G}$ and from the $\,\xi\to 0\,$
behavior (\ref{xito01}) of $F(\xi)$. One then finds for
the prefactor $F_{\alpha,n}$ the expression
\begin{equation}
F_{\alpha,n}=2\int_0^\infty\! \dd\xi\, F^n(\xi)
\qquad\quad (1-{\mbox{$\frac{1}{n}$}}<\alpha<1)
\label{defFalphan}
\end{equation}
For $0<\alpha<1-\frac{1}{n}$ the expression for $F_{\alpha,n}$ is
different and will not be needed.
Eq.\,(\ref{defFalphan}) shows that the main contribution to the sum on $r$ 
comes from $\xi\sim 1$, that is, from $r\sim
(1-z)^{-\frac{1}{\alpha}}$.
For $n=1$ and $n=2$ the integral (\ref{defFalphan}) yields the explicit
results $F_{\alpha,1}=1$, in agreement with Eq.\,(\ref{defFalpha1}),
and $F_{\alpha,2}=(\frac{1}{\alpha}-1)B_\alpha$, respectively.
For $n>\frac{1}{1-\alpha}$ the sum 
on $r$ draws its main contribution from the short distance (nonscaling)
regime $r\sim 1$, and
is of ${\cal O}(1)$ for $z\to 1$. 

By successively
substituting Eq.\,(\ref{Cn}) in Eq.\,(\ref{expGAB}), neglecting the
${\cal O}$({\bf V}$^2)$ terms in that equation -- which is justified in
Appendix B -- , then substituting Eq.\,(\ref{expGAB})
in Eq.\,(\ref{defCMM}),
expanding ${\sf G}_A(z)$ and ${\sf G}_B(z)$ according to
Eq.\,(\ref{expGA}), and using the $z\to 1$ behavior of
$\sum_r\hat{G}^n(r,z)$ obtained in Eqs.\,(\ref{fred1}) and (\ref{sumGn}) 
we find
\begin{align}
&C_{MM'}(z)=-2(1-z)^{-1}m_1m_1^\prime\nonumber\\
&\phantom{xx}-(1-z)^{\frac{1}{\alpha}-2}B_\alpha\Big(3-\mbox{$\frac{1}{\alpha}$}\Big)(m_1m_2^\prime+m_2 
m_1^\prime)\nonumber\\
&\phantom{xx}-(1-z)^{\frac{2}{\alpha}-3}\Big[\Big(4-\mbox{$\frac{2}{\alpha}$}\Big)B_\alpha^2(m_1
m_3^\prime+m_3m_1^\prime+m_2
m_2^\prime)+2F_{\alpha,3}m_2 m_2^\prime\Big]\nonumber\\
&\phantom{xx}-\ldots+{\cal O}\big(1\big)
\label{resCMM}
\end{align}
Here the dots stand for terms of order
$(1-z)^{-1+k(\frac{1}{\alpha}-1)}$, with $k=3,4,\ldots;$ 
and the $m'_n$ are related to $M'$ in the same way as the $m_n$ are to $M$.
For $t\to\infty$ we therefore find by substituting (\ref{resCMM}) in
Eq.\,(\ref{defMMav}) and Laplace inverting 
\begin{equation}
\begin{split}
&\overline{M(t)M'(t)}=m_1 m_1^\prime t^2\\
&\phantom{xx}
+\frac{B_\alpha}{\Gamma(3-\frac{1}{\alpha})}(m_1 m_2^\prime+m_2 m_1^\prime) 
t^{3-\frac{1}{\alpha}}\\
&\phantom{xx}
+\Big[\frac{B_\alpha^2}{\Gamma(4-\frac{2}{\alpha})}(m_1m_3^\prime+m_3m_1^\prime
+m_2
m_2^\prime)+\frac{2F_{\alpha,3}}{\Gamma(5-\frac{2}{\alpha})}m_2
m_2^\prime\Big]t^{4-\frac{2}{\alpha}}\\
&\phantom{xx}+\ldots +{\cal O}(t)
\end{split}\nonumber\label{moyenneMM}
\end{equation}
\vspace{-13mm}
\begin{equation}{}
\end{equation}
The successive terms in the above series all have one power of $t$ more
than the corresponding terms in the series (\ref{moyenneM}) for
$\overline{M(t)}$,
and the number of nonanalytic terms between the leading and the ${\cal
O}(t)$ term is once more equal to $n_\alpha-1$.
The product
$\overline{M(t)}$ $\overline{M'(t)}$, which follows from 
Eq.\,(\ref{moyenneM}), now has to be subtracted
from the series (\ref{moyenneMM}).
This exactly cancels the terms in (\ref{moyenneMM})
proportional to $t^2$ and to $t^{3-\frac{1}{\alpha}}$ 
but leaves those proportional to $t^{4-\frac{2}{\alpha}}$ and of ${\cal
O}(t)$. The $t^{4-\frac{2}{\alpha}}$ terms are leading
only if $\frac{2}{3}<\alpha<1$. Hence
we find for the correlation
between  observables $M$ and $M'$ in the limit $t\to\infty$
\begin{equation}
\overline{\Delta M(t)\Delta M'(t)}\simeq
{\cal B}_\alpha^2\,
m_2 m_2^\prime\;t^{4-\frac{2}{\alpha}}
\label{univ1}
\end{equation}
valid for $\frac{2}{3}<\alpha<1$, and in which
\begin{equation}
{\cal B}_\alpha^2=
\frac{2F_{\alpha,3}}{\Gamma(5-\frac{2}{\alpha})}
+\frac{B_\alpha^2}{\Gamma(4-\frac{2}{\alpha})}
-\frac{B_\alpha^2}{\Gamma^2(3-\frac{1}{\alpha})}
\label{defcalB}
\end{equation}
We have supposed here that $m_2,\,m'_2\neq 0$. The preceding
analysis 
changes when either of these two coefficients vanishes. We do not know
of any physically interesting examples where this happens, and do not
pursue our analysis in this direction.

The borderline case $\alpha=\frac{2}{3}$ is considered in
Sec.\,\ref{secspecial}. 
In the interval $0<\alpha<\frac{2}{3}$ the calculation of the present
section applies, but with the result that
\begin{equation}
\overline{\Delta M(t)\Delta M'(t)}\simeq {\kappa}_{MM'}\,t \qquad
(0<\alpha<\mbox{$\frac{2}{3}$})
\label{nonuniv}
\end{equation}
in which the coefficient ${\kappa}_{MM'}$ 
has contributions from the ${\cal O}(1)$ terms in Eq.\,(\ref{moyenneM})
and the ${\cal O}(t)$ terms in Eq.\,(\ref{moyenneMM}), and
{\it does not factor} into an $M$ and 
an $M'$ dependent
constant. This difference between Eqs.\,(\ref{univ1}) and (\ref{nonuniv})
is crucial for the phenomenon of universality discussed in
Sec.\,\ref{secuniversality}.

%%%%%%%%%%%%%%%%%%%%%%%%%%%%%%%%%%%%%%%%%%%%%%%%%%%%%%%%%%

\section{Riemann walk of exponent $\,1<\alpha<2$}
\label{secalpha12}

\subsection{Averages}
\label{seccalcM2}

The calculation of $\overline{M(t)}$ starts again from the series 
(\ref{defCM}) for $C_M(z)$. The calculation in the exponent regime
$1<\alpha<2$ is different from that of the preceding section because now
$G_0(z)$ diverges as $z\to 1$. Since $g_A(z)$ remains
finite for $z\to 1$, this suggests that we use
Eq.\,(\ref{relGg}) and expand ${\sf G}_A(z)$
in powers of $g_A(z)/G_0(z)$. This yields
\begin{equation}
C_M(z)=\sum_{A\neq\emptyset}\,\mu_A\,\frac{1}{G_0(z)}\,
\Big[\,\, 1 +\frac{{g}_A(z)}{G_0(z)} +\frac{{g}_A^2(z)}{G_0^2(z)} +
\, {\cal O}(\frac{{g}_A^3}{G_0^3})\,\, \Big] 
\label{CMseries}
\end{equation}
We now substitute in Eq.\,(\ref{CMseries}) 
the expansion 
(\ref{G3}) for $G_0(z)$ and use that $g_A(1)$ is finite. 
The result is a power series in $1-z$
in which there
appear coefficients
that we denote again by $m_n$ but that are defined 
for $1\leq\alpha<2$ as
\begin{equation}
m_n[M]=-\sum_{A\neq\emptyset}\,\mu_A\,g_A^n(1)\qquad
(n=0,1,2,\ldots;\,\,\,1\leq\alpha<2) 
\label{defmn}
\end{equation}
It will turn out that we need only the leading term, which is
\begin{equation}
C_M(z)\simeq
\left\{
\begin{array}{l}
A_\alpha^{-1}m_0(1-z)^{1-\frac{1}{\alpha}} \qquad(m_0\neq0)\\[3mm]
A_\alpha^{-2}m_1(1-z)^{2-\frac{2}{\alpha}} \qquad(m_0=0, m_1\neq 0)
\end{array}
\right.
\label{serCM}
\end{equation}
We pause to note that in the terminology of Sec.\,\ref{secobservables}
the condition $m_0\neq 0$ characterizes the bulk or {\it "S-like"}
observables,
and the condition $m_0=0$ the surface or {\it "I-like"}
observables. This is the
first equation where a difference appears between these two
subclasses; in its analog, Eq.\,(\ref{threefive}) of the preceding
section, no such distinction appears.

Upon using Eq.\,(\ref{serCM}) in Eq.\,(\ref{defMav}) 
we obtain after an inverse Laplace transformation the asymptotic
expansion of $\overline{M(t)}$ as $t\to\infty$, 
\begin{equation}
\overline{M(t)}\simeq
\left\{
\begin{array}{l}
[A_\alpha\Gamma(1+\frac{1}{\alpha})]^{-1}m_0\,t^{\frac{1}{\alpha}} 
\qquad(m_0\neq0)\\[3mm]
[A_\alpha\Gamma(\frac{2}{\alpha})]^{-1}m_1\,t^{\frac{2}{\alpha}-1} 
\qquad(m_0=0, m_1\neq 0)
\end{array}
\right.
\label{serM} 
\end{equation}
where the dots indicate terms of lower order in $t$.

\subsection{Correlations}
\label{seccalcMM2}

For the calculation of the correlation $\overline{M(t)M'(t)}$ 
via Eqs.\,(\ref{defMMav}) and (\ref{defCMM}) we have to return 
again to expression 
(\ref{decompGAB}) for $1/{\sf G}_{A\cup (r+B)}(z)$, which is needed
in Eq.\,(\ref{defCMM}). We use 
Eqs.\,(\ref{decompGAB}), (\ref{relGg}), and (\ref{deff}) to rewrite this quantity as
\begin{equation}
\frac{1}{{\sf G}_{A\cup (r+B)}(z)}=\frac{1}{G_0(z)}\,
\frac{2(1-f(r,z))-\frac{{g}_A(z)}{G_0(z)}-\frac{{g}_B(z)}{G_0(z)}}
{1-f^2(r,z)-\frac{{g}_A(z)}{G_0(z)}-\frac{{g}_B(z)}{G_0(z)}
+\frac{{g}_A(z){g}_B(z)}{G_0^2(z)}} + {\cal O}({\text{\bf V}}^2) 
\label{decompGAB2}
\end{equation} 
The function $f(r,z)$ was defined in Eq.\,(\ref{deff}). 
In the scaling limit $f(r,z),$\,\, ${g}_A(z)$, and ${g}_B(z)$
have finite limits, whereas $G_0(z)$ diverges. An expansion in inverse
powers of $G_0(z)$ corresponds therefore to an expansion in ascending
powers of $1-z$. Writing for short $f,\, {g}_A,\, {g}_B$, and $G_0$  when
$f(r,z),$\, ${g}_A(z),$\, ${g}_B(z)$, and $G_0(z)$ are meant, we find
after a straightforward calculation 
\begin{eqnarray}
C_{MM'}(z)&=&
\sum_{r=-\infty}^\infty\frac{1}{G_0} \frac{f}{1+f}
\sum_{A,B\neq\emptyset}\mu_A\mu'_B
\Big[ \,\,2\, + (2+f)\,\frac{{g}_A+{g}_B}{G_0}\nonumber\\
&&\phantom{x}+\frac{2-2f-f^3}{(1-f)(1+f)^2}\,
\frac{{g}_A^2+{g}_B^2}{G_0^2}\nonumber\\
&&\phantom{x}+\frac{2}{(1-f)(1+f)^2}\,\,\frac{{g}_A{g}_B}{G_0^2}
\,+\, {\cal O}(\frac{{g}_A^3}{G_0^3},\frac{{g}_B^3}{G_0^3})
\,\,\Big]
\label{serCMM}
\end{eqnarray}
Let us write
$m'_n=m_n[M']$ for the coefficients that characterize
the observable $M'$. 
The two distinct cases described by Eq.\,(\ref{serM}) now lead to
the following possibilities. 

{\it Case} (i): $m_0\neq0$ and $m'_0\neq 0$. 
In this case the leading term in
the expression in brackets in Eq.\,(\ref{serCMM}) survives under the sum
on $A$ and $B$. 

{\it Case} (ii): $m_0=0,\, m_1\neq 0$, and $m'_0\neq 0$. In this case 
in order to survive a 
term in the bracketed expression should contain at least one factor
${g}_A(z)$. 

{\it Case} (iii): $m_0=m'_0=0$ but $m_1\neq 0$ and $m'_1\neq 0$. In this
case a term in order 
to survive must contain at least one factor ${g}_A(z)$ and one factor
${g}_B(z)$.\\
Upon using in each of these cases for $G_0(z)$the expansion (\ref{G3}),
passing to the scaling limit, and writing $f$ for $f(\xi)$,
we find that the result is
\begin{equation}
C_{MM'}(z)\simeq\left\{
\begin{array}{l}
-A_\alpha^{-1}(1-z)^{1-\frac{2}{\alpha}}f_{00}\,m_0m'_0
\label{serCMM00}\\[2mm]
-A_\alpha^{-1}(1-z)^{2-\frac{3}{\alpha}}f_{10}\,m_1m'_0
\label{serCMM10}\\[2mm]
-A_\alpha^{-1}(1-z)^{3-\frac{4}{\alpha}}f_{11}\,m_1m'_1
\label{serCMM11}
\end{array}
\right.
\end{equation}
in the three cases {(i), (ii),\,} and {(iii),\,} respectively; 
here the coefficients $f_{k\ell}$ represent the integrals
\begin{eqnarray}
f_{00}&=&4\int_0^\infty\dd\xi \,\,f(1+f)^{-1}\nonumber\\[2mm]
f_{10}&=&2\int_0^\infty \dd\xi \,\,f(2+f)(1+f)^{-1}
\label{deff00}\\[2mm]
f_{11}&=&4\int_0^\infty \dd\xi \,\,f(1-f)^{-1}(1+f)^{-3}\nonumber
\end{eqnarray}
After substituting Eqs.\,(\ref{serCMM11}) in Eq.\,(\ref{defMMav}) and
carrying out the inverse Laplace transformation
we find, in the limit $t\to\infty$, 
\begin{equation}
\overline{M(t)M'(t)}\simeq
\left\{
\begin{array}{rl}
\Gamma^{-1}(
1+\frac{2}{\alpha})&\!\!A_\alpha^{-1}f_{00}\,m_0m'_0\,\,t^{\frac{2}{\alpha}}\\[3mm]
\Gamma^{-1}(
\frac{3}{\alpha})&\!\!A_\alpha^{-2}f_{10}\,m_1m'_0\,\,t^{\frac{3}{\alpha}-1}\\[3mm]
\Gamma^{-1}(-1+\frac{4}{\alpha})&\!\!A_\alpha^{-3}f_{11}\,m_1m'_1\,\,t^{\frac{4}
{\alpha}-2}
\end{array}
\right.
\label{resMM}
\end{equation}
respectively, for the three cases distinguished above.
Upon combining these results with those of Section \ref{seccalcM2} one
obtains, for $t\to\infty$,
\begin{equation}
\overline{\Delta M(t)\Delta M'(t)}\simeq
\left\{
\begin{array}{l}
{\cal B}_{\alpha 00}^2\, m_0m'_0\,t^{\frac{2}{\alpha}}\\[3mm]
{\cal B}_{\alpha 10}^2\, m_1m'_0\,t^{\frac{3}{\alpha}-1}\\[3mm]
{\cal B}_{\alpha 11}^2\, m_1m'_1\,t^{\frac{4}{\alpha}-2}
\end{array}
\right.
\label{rescorr}
\end{equation}
in which the coefficients ${\cal B}_{\alpha k\ell}$ are given by
\begin{equation}
\begin{array}{l}
{\cal B}_{\alpha 00}^2=A_\alpha^{-1}[f_{00}\Gamma^{-1}(1+\frac{2}{\alpha})-
A_\alpha^{-1}\Gamma^{-2}(1+\frac{1}{\alpha})]\\[3mm]
{\cal B}_{\alpha 10}^2=A_\alpha^{-2}[f_{10}\Gamma^{-1}(\frac{3}{\alpha})-
A_\alpha^{-1}\Gamma^{-1}(1+\frac{1}{\alpha})\Gamma^{-1}(\frac{2}{\alpha})]\\[3mm
]
{\cal B}_{\alpha 11}^2=A_\alpha^{-3}[f_{11}\Gamma^{-1}(-1+\frac{4}{\alpha})-
A_\alpha^{-1}\Gamma^{-2}(\frac{2}{\alpha})]
\end{array}
\label{defckl}
\end{equation}
in the three cases {(i),\, (ii),\,} and {(iii)\,} defined above,
respectively.
We recall that the $f_{k\ell}$ on the RHS of Eq.\,(\ref{defckl}) are
given by Eq.\,(\ref{deff00}) as integrals on $f(\xi)$, with $f(\xi)$
in turn given by Eq.\,(\ref{defF}).

%%%%%%%%%%%%%%%%%%%%%%%%%%%%%%%%%%%%%%%%%%%%%%%%%%%%%%%%%%%%%%%%%%%%%%%%%%

\section{Riemann walk of exponents $\alpha=\frac{2}{3}$ and $\alpha=1$}
\label{secspecial}

\subsection{Exponent $\alpha=\frac{2}{3}$}

In this special case $\overline{M(t)}$ is still given by
Eq.\,(\ref{threefive}). 
However, the calculation of $\overline{M(t)M'(t)}$ has to
be reconsidered, as signalled by the fact that $F_{\alpha,3}$ in
Eq.\,(\ref{resCMM}) diverges for $\alpha\to\frac{2}{3}^+$. In order to
calculate $\sum_r\hat{G}(r,z)$ we cannot now use Eq.\,(\ref{sumG3}). Instead
we replace $\hat{G}(r,z)$ by its scaling form (\ref{G3}) but take into
account that $|\xi|$ has a lower cutoff
$\,|\xi|\sim\mbox{cst}\!\times\!(1-z)^{\frac{2}{3}}$. This gives
\begin{eqnarray}
\sum_r\hat{G}^3(r,z)&\simeq&2\int_{\mbox{\small{cst}}
\times(1-z)^{2/3}}^\infty
\dd\xi\,F^3(\xi)\nonumber\\
&\simeq& \mbox{$\frac{1}{2}$}{\cal C}^2\,\log(1-z)^{-1}
+{\cal O}(1)\qquad (z\to 1)
\label{sum3}
\end{eqnarray}
where in the second step we used Eq.\,(\ref{defA}) and found for the
coefficient the value
${\cal C}^2=2^{12}3^{-11/2} \pi^{-3}\zeta^3(\frac{3}{2})$.
In this case, due to the $\log(1-z)$ in
the equation above, $\overline{M(t)M'(t)}$ is larger than
the product $\overline{M(t)}$\,\,$\overline{M'(t)}$ by a factor $\log t$, and
determines by itself alone the final result, which reads
\begin{equation}
\overline{\Delta M(t)\Delta M'(t)}\,\simeq\,{\cal C}^2\,m_2m'_2\,\,t\log t \qquad (t\to\infty)
\label{univ23}
\end{equation}
This $t\log t$ behavior is the same as in the well-known case of the
simple random walk in spatial dimension $d=3$ \cite{WH}.

\subsection{Exponent $\alpha=1$}

The case of L\'evy exponent $\alpha=1$ is subtler than the
others. Since it is closely analogous to the simple
random walk in dimension $d=2$ \cite{WCH}, 
we will not present
all steps in detail. Eq.\,(\ref{G2.5}) shows that for $\alpha=1$ the Green
function in the origin, $G_0(z)$, diverges as $z\to 1$.
We can therefore expand $C_M(z)$ as a
series in the
same way as in Eq.\,(\ref{CMseries}). Since here again
$g_A(z)=g_A(1)+{\cal O}(1-z)$, and in view of the logarithmic behavior
(\ref{G2.5}), this series now leads to an expansion of $C_M(z)$
in inverse powers of 
$G_0(z)$. If the first nonzero term is of order $k+1$,
then we have explicitly
\begin{equation}
C_M(z)=-m_kG_0^{-k-1}(z)-m_{k+1}G_0^{-k-2}(z)-m_{k+2}G_0^{-k-3}(z)+\ldots
\label{CMserieslog}
\end{equation} 
with the $m_n$ defined by Eq.\,(\ref{defmn}).  
The cases of physical interest have $k=0$ (bulk observables) 
or $k=1$ (surface observables), but it will be notationally convenient
to keep $k$ as a parameter.
We will also refer to it as the {\it order} of $M$.

To find $C_{MM'}(z)$ we may still start from 
Eq.\,(\ref{decompGAB2}), but now the
expansion of this equation runs differently. The reason is that for
$z\to 1$ at fixed $\xi$ the function $f(r,z)$ (defined by (\ref{deff}))
behaves as $F(\xi)/G_0(z)$
and so is of the same order as $g_A(z)/G_0(z)$ and 
$g_B(z)/G_0(z)$. We therefore have to perform a double expansion of the
RHS of Eq.\,(\ref{decompGAB2}) in terms of on the one hand $F/G_0$ and on the
other hand $g_A/G_0$ and $g_B/G_0$. The sum on $r$, which in the scaling
limit becomes an integral on $\xi$, then leads 
to the appearance of coefficients $F_{1,n}$ defined as in
Eq.\,(\ref{defFalphan})
but with $\alpha=1$. 
Special cases are $F_{1,1}=1$ and $F_{1,2}=\frac{1}{3}$.
Let $m_k$ and $m'_{k'}$ be the first nonzero coefficients in the
expansions of $C_M(z)$ and $C_{M'}(z)$, respectively. Then we find for
$C_{MM'}(z)$, retaining only the three leading order terms in the limit
$z\to 1$,   
\begin{align}
C_{MM'}(z)\simeq&-\frac{1}{(1-z)G_0^{k+k'+2}(z)}\Big[
2m_km_{k'}^\prime\nonumber\\
-&G_0^{-1}(z)\Big(\mbox{$\frac{1}{3}$}(k+k'+2)a_2m_km_{k'}^{\prime}-
2(m_km^\prime_{k'+1}+m_{k+1}m_{k'}^\prime)\Big)\nonumber\\
+&G_0^{-2}(z)\Big(2F_{1,3}(k+1)(k'+1)a_3 m_k m_{k'}^\prime\nonumber\\
&\phantom{G_0^{-2}(z)\Big(}-\mbox{$\frac{1}{3}$}(k+k'+3)a_2(m_km_{k'+1}^\prime+
m_{k+1}m_{k'}^\prime)\nonumber\\  
&\phantom{G_0^{-2}(z)\Big(}+2(m_km_{k'+2}^\prime+m_{k+1}m_{k'+1}^\prime+
m_{k+2}m_{k'}^\prime)\Big)\Big]
%\nonumber\\
%&+\,\cdots\,\Big]
\label{CMMserieslog}
\end{align} 
The inverse Laplace transforms of $C_M(z)$ 
and $C_{MM'}(z)$ may be found with the help of the explicit expression
(\ref{G2.5}) for $G_0(z)$ and
the integrals of Ref.\,\cite{WCH}. 
We state only the explicit result for $\overline{M(t)}$, which is,
for $t\to\infty$,
\begin{align}
\overline{M(t)}\simeq\frac{3^{k+1}t}{\log^{k+1}ct}\Big[&m_k
\,+\,\frac{1}{\log ct}\Big((1-\gamma)(k+1)m_k+3m_{k+1}\Big)\nonumber\\
&+\,\frac{1}{\log^2 ct}\Big(
(1-{\mbox{$\frac{1}{12}$}}\pi^2-\gamma
+\mbox{$\frac{1}{2}$}\gamma^2)(k+1)(k+2)m_k\nonumber\\
&\phantom{XXXXX}-3(1-\gamma)(k+2)m_{k+1}+9m_{k+2}
\Big)
\Big]
\end{align}
in which $\gamma=0.577215...$ denotes Euler's constant.
Both $\overline{M(t)M'(t)}$
and the product $\overline{M(t)}$\,\,$\overline{M'(t)}$ the appear as
$t^2$ times a
power series in $1/\log ct$ of which the leading term is of order $k+k'+2$,
and in which the three leading orders have to be retained. Upon carrying
out the subtraction one finds that
the two leading orders cancel and the correlation
$\overline{\Delta M(t)\Delta M'(t)}$ 
appears to be proportional to $t^2/\log^{k+k'+4} ct$.
Explicitly, as $t\to\infty$,
\begin{equation}
\overline{\Delta M(t)\Delta M'(t)}\simeq {\cal A}^2 (k+1)(k'+1)
m_km'_{k'}\frac{3^{k+k'+2}\,t^2}{\log^{k+k'+4} ct}
\qquad (t\to\infty)
\label{corrlog}
\end{equation}
in which the coefficient ${\cal A}$ is given by
\begin{equation}
{\cal A}^2=1\,+\,(F_{1,3}-\mbox{$\frac{1}{6}$})\pi^2
\end{equation}
Numerical evaluation gives $F_{1,3}=0.27415...$, whence
${\cal A}^2=2.0608...$. 
Eq.\,(\ref{corrlog}) is the same as 
for the two-dimensional
simple random walk \cite{Torney,Weiss,Hughes,WCH}, 
but with a different constant ${\cal A}$. 

%%%%%%%%%%%%%%%%%%%%%%%%%%%%%%%%%%%%%%%%%%%%%%%%%%%%%%%%%%%%%%%%%%%%%%%%%%%%

\section{Universality of fluctuations}
\label{secuniversality}

We consider in this section the 
normalized deviations from average
\begin{equation}
\theta_M(t)=
\frac{\Delta M(t)}{{\overline{\Delta M^2(t)}}^{1/2}}
\label{defxiM}
\end{equation}
These random functions of time satisfy by construction 
\begin{equation}
\overline{\theta_M(t)}=0, \qquad \overline{\theta_M^2(t)}=1
\label{avxiM}
\end{equation}
We consider now two arbitrary observables $M$ and $M'$.  
When $\frac{2}{3}\leq\alpha<1$ we have 
from Eq.\,(\ref{defxiM}) together with either Eq.\,(\ref{univ23}) or
Eq.\,(\ref{univ1}) that
\begin{equation}
\overline{\theta_M(t)\theta_{M'}(t)}=1  
\label{xiMxiM}
\end{equation}
It then follows from Eqs.\,(\ref{avxiM}) and (\ref{xiMxiM}) that the
difference $\theta_M-\theta_{M'}$ is a random variable of zero
average and zero variance. Such a random variable can only be itself
equal to zero. We therefore deduce
%along the same lines as in Refs.\,\cite{WCH,WH} 
that, when $\frac{2}{3}\leq\alpha<1$, 
in the limit $t\to\infty$ all $\theta_M(t)$
are equal to a single random variable, which 
we will call $\Theta_\alpha(t)$, thus indicating explicitly its $\alpha$
dependence. 

When $1\leq\alpha<2$ we have for two observables $M$ and $M'$ {\it whose
orders, $k$ and $k'$, are equal} from Eq.\,(\ref{defxiM}) and either
Eq.\,(\ref{corrlog}) or 
Eq.\,(\ref{rescorr}) again the result (\ref{xiMxiM}).
Hence, when $1\leq\alpha<2$,
in the limit $t\to\infty$ all $\theta_M(t)$ with $k=0$
are equal to a single random variable -- that we will call
$\Theta_{\alpha 0}(t)$ --,
and all $\theta_M(t)$ with
$k=1$ are similarly
equal to a single random variable -- that we will call
$\Theta_{\alpha 1}(t)$. 
The variables $\Theta_\alpha(t)$,\,
$\Theta_{\alpha 0}(t)$, and $\Theta_{\alpha 1}(t)$ 
are {\it universal} in the sense
that they are independent of the observables $M$ (but depend at most on
their order).

In each of these cases the 
key ingredient necessary for arriving at Eq.\,(\ref{xiMxiM})
is the factorization 
of $\overline{M(t)M'(t)}$ into an $M$ and an $M'$
dependent part. This also explains why for $\alpha<\frac{2}{3}$ the same
reasoning fails.

The {\it cross correlation} between $\Theta_{\alpha 0}(t)$ 
and $\Theta_{\alpha 1}(t)$ is easily
found from the correlation between a $\theta_M(t)$ and a $\theta_{M'}(t)$ with
$k=0$ and $k'=1$, and use of the second one of Eqs.\,(\ref{rescorr}).
The answer is independent of the choice of $M$ and $M'$, as it had to
be, and reads
\begin{equation}
\overline{\Theta_{\alpha 0}(t)\Theta_{\alpha 1}(t)}
={\cal B}_{\alpha 10}^2 ({\cal B}_{\alpha
00}{\cal B}_{\alpha 11})^{-1}
\label{xi0xi1}
\end{equation}
The coefficient ratio on the RHS of this equation
depends only on the exponent $\alpha$ and must
necessarily be less than unity. In the limit $\alpha\to 1^+$ it
approaches unity and for $\alpha<1$
the distinction between surface and bulk
observables is no longer reflected in the fluctuations.

Upon combining all these conclusions we get explicitly
\begin{equation}
\Delta M(t)=
\left\{
\begin{array}{ll}
m_2\,{\cal C}\,(t\log t)^{\frac{1}{2}}\,\Theta_{\frac{2}{3}}(t)
\qquad & (\alpha=\mbox{$\frac{2}{3}$})\\[2mm]
m_2\,{\cal B}_\alpha\,t^{2-\frac{1}{\alpha}}\,\Theta_\alpha(t)
\qquad &(\mbox{$\frac{2}{3}$}<\alpha<1)\\[2mm]
m_k\,{\cal A}\,(k+1)\,t\,(\mbox{$\frac{1}{3}$}\log ct)^{-k-2}\,\Theta_1(t)
\quad &(\alpha=1; k=0,1)\\[2mm]
m_0\,{\cal B}_{\alpha 00}\,t^{\frac{1}{\alpha}}\,\Theta_{\alpha 0}(t)
\qquad &(1\!<\!\alpha\!<\!2; k=0)\\[2mm]
m_1\,{\cal B}_{\alpha 11}\,t^{\frac{2}{\alpha}-1}\,\Theta_{\alpha 1}(t)
\qquad &(1\!<\!\alpha\!<\!2; k=1)
\end{array}
\right.
\label{relDMxi}
\end{equation}
Here all $M$ dependence is contained in the coefficients $m_n$. 

%%%%%%%%%%%%%%%%%%%%%%%%%%%%%%%%%%%%%%%%%%%%%%%%%%%%%%%%%%%%%%%%%%%%%%%%%

\section{Conclusions}
\label{secconclusions}

We have studied a large class of properties $M(t)$ of the support of the
one-dimensional $t$ step Riemann walk. These include the number $S(t)$
of distinct sites visited, and the number $I(t)$ of sequences of visited
sites. The $M(t)$ fall into two classes, the
{\it bulk} or {\it S-like} properties, and the {\it surface} or 
{\it I-like} properties.  
The asymptotic laws found in the preceding sections for the averages,
variances, and correlation of $S(t)$ and $I(t)$ have been summarized
in Table I in the Introduction.

It appears from that table that in the exponent regime
$0<\alpha\leq 1$ the ratios 
$\overline{\Delta S^2(t)}^{1/2}\!/\,\overline{S(t)}$ and 
$\overline{\Delta I^2(t)}^{1/2}\!/\,\overline{I(t)}$  
tend to zero when
$t\to\infty$, which indicates 
that the distributions of $S(t)$ and of $I(t)$
become infinitely narrowly peaked around their average. 
Hence in this exponent regime the ratio
$s(t)\equiv\overline{S(t)}/\overline{I(t)}$ 
represents the {\it average number of sites per visited sequence}. 
When $\alpha$ is strictly less than unity we have explicitly
\begin{equation}
\lim_{t\to\infty}s(t)\,=\,\frac{m_1[S]}{m_1[I]}\,=\,
\frac{\hat{G}(0,1)-\hat{G}(1,1)}{\hat{G}(0,1)+\hat{G}(1,1)}
\qquad (0<\alpha<1)
\label{exprst}
\end{equation}
The first equality is based on Eq.\,(\ref{threefive}) and in the
second one we used the definition (\ref{defmnM}) of the $m_n[M]$ and
the remarks at the end of Sec.\,\ref{secgeneralities}. 
The finiteness of the result (\ref{exprst}) 
means that in the large $t$ limit every
new step of the walk creates a new
visited sequence with a finite nonzero probability. 
This explains that in this exponent regime
the asymptotic power laws do not distinguish between
bulk and surface properties.
For $\alpha\to 1^-$ expression (\ref{exprst}) diverges, and
when $\alpha=1$ the ratio $s(t)$ increases logarithmically with $t$.

In the exponent regime $1<\alpha<2$ the appropriately scaled 
distributions of $S(t)$ and $I(t)$ are of finite width even in the limit
$t\to\infty$. The support has an "interior", bulk and
surface properties have different asymptotic power laws, and
$s(t)\to\infty$ as $t\to\infty$. In the terminology of critical
phenomena, this regime is fluctuation dominated. 
In this regime the universality of fluctuations holds in a slightly
weaker but at least as interesting a sense as for $0<\alpha\leq 1$.
To describe the fluctuations, not a single but {\it two} 
universal stochastic variables are
needed, one applying to the bulk and the other to the surface properties. 
These two variables become fully correlated in the limit $\alpha\to
1^+$.

Finally we remark that when $2/\alpha$ is equal to one of the integers
$2,3,4,\ldots$, the asymptotic laws of Table I coincide with the ones 
known to hold for the simple random walk on a lattice of dimension
$d=2/\alpha$. Similarly, the universality properties for those $\alpha$
values have their analogs in the $d$-dimensional simple random walk. 
Hence the "rule of the effective dimensionality", which states the
correspondence $\alpha\Leftrightarrow 2/d$, applies to all properties
that we have studied. Of course it must break down when the comparison
between the Riemann walk and the simple random walk is refined
sufficiently. 
Also, we have not considered the borderline value $\alpha=2$, which is 
special \cite{GillisWeiss}, and for which this rule fails.

\section*{Acknowledgments}

The authors acknowledge support from the French-Brazilian scientific
cooperation project CAPES/COFECUB 229/97.

%%%%%%%%%%%%%%%%%%%%%%%%%%%%%%%%%%%%%%%%%%%%%%%%%%%%%%%%%%%%%%%%%%%%%%%%%%%%%%%

\appendix
\section{Relations for the inverse sums ${\sf G}_A$ and $g_A$}

We collect here some elementary matrix algebra relations 
useful for dealing with the inverse sums
${\sf G}_A$ and $g_A$
occurring in the main text. The $z$ dependence of these quantities plays
no role. The presentation and notation are independent of
the body of the paper.

Let $L$ be an invertible $\ell\times\ell$ matrix. We define
the "inverse sum" 
${\cal I}(L)$ by
\begin{equation}
{\cal I}^{-1}(L)=\sum_{i,j}L_{ij}^{-1}
\label{defI}
\end{equation}  
In the remainder $\alpha, \beta,$ and $\gamma$ will denote constants.\\ 

\noindent
{\sc Property 1.} 
{\it Let $J$ be the $\ell\times\ell$ matrix with all $J_{ij}=1$, and let
$M$ be an invertible $\ell\times\ell$ matrix. Let $L=\alpha J+\gamma M$.
Then}
\begin{equation}
{\cal I}(L) = \alpha + \gamma{\cal I}(M)
\label{property1}
\end{equation}
The proof of this relation is given in Ref.\,\cite{WCH}.\\ 

\noindent
{\sc Property 2.}
{\it Let $M$ and $N$ be invertible matrices of dimensions $m\times
m$ and $n\times n$, respectively, and let $L$ be the block diagonal 
$\ell\times\ell$ matrix with blocks $M$ and $N$. 
Then}
\begin{equation}
\frac{1}{{\cal I}(L)}=\frac{1}{{\cal I}(M)} + \frac{1}{{\cal I}(N)}
\end{equation}
This follows directly from the definition (\ref{defI}).
The calculation of ${\cal I}(L)$ for an $\ell\times\ell$ matrix may be
reduced to an inversion problem of dimension less than $\ell$ 
also in certain cases where $L$ is not block
diagonal, as shown below.\\

\noindent
{\sc Property 3.}
{\it Let $J^{mn}$ be the $m\times n$ matrix with all elements equal
to 1. Let $L$ be $\ell\times\ell$ and of the form
\begin{equation}
L=
\left(
\begin{array}{cc}
\gamma M&\beta J^{mn}\\      
\beta J^{nm} & \gamma N
\end{array}
\right)
\label{Lprop3}
\end{equation}
Then}
\begin{equation}
{\cal I}(L)=\frac{\gamma^2{\cal I}(M){\cal I}(N) - \beta^2}
{\gamma{\cal I}(M) + \gamma{\cal I}(N) - 2\beta}
\label{property3}
\end{equation} 
To prove this we rewrite $L$ as $L=\beta J + \tilde{L}$, where $J$ is as
before and where
\begin{equation}
\tilde{L}=\left(
\begin{array}{cc}
\gamma M-\beta J^{mm} & 0\\     0 & \gamma N-\beta J^{nn}
\end{array}
\right)
\end{equation}
From {\sc Property 1} we have that ${\cal I}(L)=\beta + {\cal
I}(\tilde{L})$, after 
which by applying {\sc Property 2} and once more {\sc Property 1}, 
we obtain after some rearrangement
Eq.\,(\ref{property3}).
For $\beta=0$ Eq.\,(\ref{property3}) reduces to {\sc Property 2}.

In this work the need for {\sc Properties} 1 and 3 arises when the limit
$\gamma\to 0$ has to be taken. For $\gamma=0$ the matrices $L$ that
occur on the
LHS of Eqs.\,(\ref{property1}) and (\ref{property3})
are no longer invertible, but these
properties allow nevertheless ${\cal I}(L)$ to be calculated in that limit.
 
\section{Corrections to scaling}

In Eq.\,(\ref{expGAB}) we have neglected the ${\cal O}$({\bf V}$^2$) terms that
appear in Eq.\,(\ref{decompGAB}). 
Since in the last step that led to Eq.\,(\ref{univ1})
the leading order in $1-z$ went down due to
cancellations, we must now check that the ${\cal
O}$({\bf V}$^2$) terms remain subdominant. 
In this Appendix we will write Eq.\,(\ref{splitGAB}) 
in the simplified notation
$\mbox{\bf{G}}_{A\cup(r+B)}=\mbox{\bf{G}}+\mbox{\bf{W}}$
where {\bf G} and {\bf W} are the first and second matrix, respectively,
on the RHS of Eq.\,(\ref{splitGAB}). Upon writing the inverse
{\bf G}$_{A\cup(r+B)}^{-1}$ as a perturbation series in {\bf W} and
applying Eq.\,(\ref{defGA}) one finds
\begin{equation}
{\sf{G}}_{A\cup(r+B)}^{-1}=\sum_{\ell=0}^\infty
(-1)^{\ell}\sum_{c,c'}
\,[\mbox{\bf{G}}^{-1}(\mbox{\bf{WG}}^{-1})^{\ell}]_{cc'}
\label{pertseries}
\end{equation}
The $\ell=0$ term of this series is the term shown explicitly on the RHS
of Eq.\,(\ref{decompGAB}), and
has been the object of study in
Sec.\,\ref{seccorrelations1}. 
We will show here that the terms with $\ell\geq 1$
produce, in the scaling limit, only higher order corrections to the
final result. To this end we first consider $C_{MM'}(z)$ defined by
Eq.\,(\ref{defCMM}).
Let $R_\ell(z)$ denote the contribution to $C_{MM'}(z)$ from the
$\ell$th term in Eq.\,(\ref{pertseries}).
In order to estimate the order in $\,1-z\,$ of $R_\ell(z)$ as $z\to 1$
we first deduce from Eqs.\,(\ref{defV}) and (\ref{Grzscaling}) 
that in the scaling limit the matrix elements of {\bf W} behave
as $V_{a,r+b}\simeq(1-z)^{\frac{2}{\alpha}-1}
(b-a)\,F'(\xi)$, and that summing on $r$ amounts to applying
$(1-z)^{-\frac{1}{\alpha}}\int\dd\xi$. This yields the asymptotic
proportionality 
\begin{equation}
R_{\ell}(z)\sim(1-z)^{-\frac{1}{\alpha}}\int\dd\xi\,\,\frac{1}{G_0(z)}\,
\Big[\frac{(1-z)^{\frac{2}{\alpha}-1}F'(\xi)}{G_0(z)}\Big]^\ell
\label{asptprop}
\end{equation}
where $G_0(z)$ represents the order in $\,1-z\,$ of the matrix {\bf G}.
When $\ell$ is odd, this integral vanishes by symmetry, which shows that
the leading correction is of order {\bf V}$^2$, as anticipated.
For $\xi\to 0$ we have, in virtue of Eq.\,(\ref{xito01}), that
$F'(\xi)\sim\xi^{-2+\alpha}$. Hence the $\xi$ integral in
Eq.\,(\ref{asptprop}) diverges in the origin for all $\ell\geq 1$ when
$0<\alpha<1$. This signals that the main contribution comes
from $r$ values near the origin. The order in $\,1-z\,$ of $R_\ell(z)$
may then be estimated by introducing in the integral the cutoff
$|\xi|\sim\mbox{cst}\times (1-z)^{\frac{1}{\alpha}}$, which leads to
$R_\ell(z)\sim(1-z)^{0}$. 
When $\ell\geq 1$ these additive corrections to $C_{MM'}(z)$ in
Eq.\,(\ref{resCMM}) are negligible, therefore, with respect to
the $(1-z)^{\frac{2}{\alpha}-3}$ term which, in the relevant exponent
regime $\frac{2}{3}<\alpha<1$,
determines the final result.

%%%%%%%%%%%%%%%%%%%%%%%%%%%%%%%%%%%%%%%%%%%%%%%%%%%%%%%%%%%%%%%%%%%%%%%%%%%%%

\end{document}